\documentclass[a4paper,fleqn,usenatbib]{mnras}
\usepackage{natbib}
\usepackage{epsf}

\usepackage[T1]{fontenc}
\usepackage{ae,aecompl}

\usepackage{graphicx}	% Including figure files
\usepackage{amsmath}	% Advanced maths commands
\usepackage{amssymb}	% Extra maths symbols

\title[Polarized synchrotron emission in quiescent BHs]{Polarized synchrotron emission in quiescent black hole X-ray transients}

\author[D. M. Russell et al.]
{David M. Russell$^{1}$\thanks{E-mail: dave.russell@nyu.edu}, Tariq Shahbaz$^{2,3}$, Fraser Lewis$^{4,5}$ and Elena Gallo$^{6}$
\\
$^1$New York University Abu Dhabi, PO Box 129188, Abu Dhabi, UAE \\
$^2$Instituto de Astrof\'isica de Canarias (IAC), E-38200 La Laguna, Tenerife, Spain \\
$^3$Departamento de Astrof\'isica, Universidad de La Laguna (ULL), E-38206 La Laguna, Tenerife, Spain \\
$^4$Faulkes Telescope Project, School of Physics, and Astronomy, Cardiff University, The Parade, Cardiff, CF24 3AA, Wales \\
$^5$Astrophysics Research Institute, Liverpool John Moores University, 146 Brownlow Hill, Liverpool L3 5RF, UK  \\
$^6$Department of Astronomy, University of Michigan, Ann Arbor, MI 48109-1042, USA
}

\date{Accepted XXX. Received YYY; in original form ZZZ}

\pubyear{2016}

\def\simlt{\mathrel{\rlap{\lower 3pt\hbox{$\sim$}}
        \raise 2.0pt\hbox{$<$}}}
\def\simgt{\mathrel{\rlap{\lower 3pt\hbox{$\sim$}}
        \raise 2.0pt\hbox{$>$}}}

\begin{document}
\label{firstpage}
\pagerange{\pageref{firstpage}--\pageref{lastpage}}
\maketitle

\begin{abstract}
We present near-infrared polarimetric observations of the black hole X-ray binaries Swift J1357.2--0933 and A0620--00. In both sources, recent studies have demonstrated the presence of variable infrared synchrotron emission in quiescence, most likely from weak compact jets. For Swift J1357.2--0933 we find that the synchrotron emission is polarized at a level of $8.0 \pm 2.5$ per cent (a 3.2 $\sigma$ detection of intrinsic polarization). The mean magnitude and rms variability of the flux (fractional rms of 19--24 per cent in $K_{\rm S}$-band) agree with previous observations. These properties imply a continuously launched (stable on long timescales), highly variable (on short timescales) jet in the Swift J1357.2--0933 system in quiescence, which has a moderately tangled magnetic field close to the base of the jet.
We find that for A0620--00, there are likely to be three components to the optical--infrared polarization; interstellar dust along the line of sight, scattering within the system, and an additional source that changes the polarization position angle in the reddest ($H$ and $K_{\rm S}$) wave-bands. We interpret this as a stronger contribution of synchrotron emission, and by subtracting the line-of-sight polarization, we measure an excess of $\sim 1.25 \pm 0.28$ per cent polarization and a position angle of the magnetic field vector that is consistent with being parallel with the axis of the resolved radio jet. These results imply that weak jets in low luminosity accreting systems have magnetic fields which possess similarly tangled fields compared to the more luminous, hard state jets in X-ray binaries.
\end{abstract}


\section{Introduction}

Most low-mass X-ray binaries (LMXBs) spend the majority of their time in a state of quiescence, accreting at very low luminosities; $\sim 10^{30}$--$10^{\sim 33}$ erg s$^{-1}$ \citep*[e.g.][]{vanpet87,konget02,homaet13,plotet13}. Studies of quiescent LMXBs have provided black hole \cite[e.g.][]{casajo14} and neutron star \cite*[e.g.][]{davaet05,munoet09,lattet14,shahet14} mass estimates, possible evidence for event horizons (\citealt{garcet01}, although see \citealt*{fendet03}), constraints on the equation of state of neutron stars \cite[e.g.][]{heinet03,guilet13} and even constraints on braneworld gravity models and the size of extra dimensions \citep*{psalet07,joha09,gonzet14}. Evidence for accretion onto the compact object at these lowest luminosities implies these sources never truly switch off. X-ray variability \citep[e.g.][]{rutlet01,degewi12,bernca14,cotiet14} and the UV/optical/infrared properties \citep[e.g.][]{oroset94,zuriet03,shahet05,cantet08,yanget12,macdet14,bernet16,kollet16} in quiescence can be explained by a truncated disc and a radiatively inefficient accretion flow. The X-ray and optical fluxes can be positively correlated on long \citep[e.g.][]{bernet13} and short \citep[e.g.][]{hyneet04} timescales.

Jets appear to play a relatively more energetically important role at low luminosities compared to higher luminosities in black hole LMXBs \citep*{fendet03,russet10} and jets may contribute to, or even dominate most of the spectrum from radio to X-ray in quiescence in some sources \citep*[e.g.][]{xieet14,market15,plotet15,plotet16}. Several X-ray binaries have confirmed radio detections in quiescence \citep*{gallet05,gallet06,gallet14,millet11,millet15,straet12,chomet13,dzibet15,tetaet16} and they are all black hole candidate systems. Spectral evidence for synchrotron emission in the optical--infrared (OIR) regime in quiescence has been found in some black hole LMXBs \citep*{gallet07,geliet10,fronet11,shahet13,plotet16} and one neutron star system \citep{baglet13}. 

During outbursts of black hole LMXBs, continuously launched, synchrotron-emitting jets are commonly produced during the hard spectral state \citep*[see][ for a recent review]{fendga14}. The hard state commonly occurs at the beginning and end of outbursts, from luminosities above quiescence ($> 10^{\sim 33}$ erg s$^{-1}$) up to $\sim 1$--15 per cent of the Eddington luminosity \citep[e.g.][]{bell10,kollet16}. The hard state is associated with a highly variable (on short timescales) hard power law X-ray spectrum that likely originates in an inner, radiatively inefficient accretion flow close to the black hole. Optically thin synchrotron emission is observed at OIR wavelengths and originates from the inner parts of the jets close to their launch region \citep[e.g.][ and references therein]{corbfe02,hyneet03,russet13}.

Polarimetry can be used as a tool to probe the magnetic field configuration in these inner parts of the jets, since the polarization of optically thin synchrotron emission is dependent on the level of ordering of the magnetic field. Detections of (in some cases variable) intrinsic polarization on levels of a few per cent have now been confirmed in several LMXBs. Three black hole candidates were found to have intrinsic polarization at infrared wavelengths during outburst, all during the hard state when jets are expected to be launched, at levels of $\sim 4$--7 per cent (GRO J1655--40), $\sim 2.4$ per cent (XTE J1550--564) and up to 2.8 per cent \citep[GX 339--4;][]{russfe08,chatet11,russet11}. In the latter source, variability of the polarization was seen on timescales of minutes. Just recently, during the bright outburst of V404 Cyg in 2015, flares of optical polarization were detected and interpreted as synchrotron emission from a variable jet \citep{lipuet16,shahet16}. In neutron star LMXBs, intrinsic infrared or optical polarization was detected in Sco X--1, Cyg X--2 and 4U 0614+09, at similar levels as the black hole LMXBs (up to $\sim 10$ per cent in Cyg X-2) and short timescale variations of polarization were observed in Sco X-1 \citep*{russfe08,shahet08,russet11,baglet14b}. These measurements have so far suggested that the magnetic fields near the jet base are largely tangled, chaotic and on average, aligned with the jet axis. One exception is Cyg X--1, which appears to have a highly ordered, stable magnetic field near its jet base that is perpendicular to the axis of the jet \citep{russsh14,rodret15}, and V404 Cyg which had flares of low level polarization, with the magnetic field also perpendicular to the jet axis \citep{shahet16}. These levels of polarization are similar to those often seen from LMXB jets at radio frequencies during the hard state, although higher levels (tens of per cent) of radio polarization have been detected, especially during spectral state transitions when discrete ejections are launched \citep[e.g.][]{corbet00,brocet07,brocet13,curret14}. The typical hard state infrared and radio polarization levels are also similar to the polarization measurements of synchrotron emission from the cores of active galactic nuclei (AGN), which are also polarized on levels of a few to tens of per cent \citep[e.g.][]{listho05,barret10,lopeet14}. With a larger sample of X-ray binaries and AGN, it will be possible to test for differences in the level of magnetic field ordering, and orientation, near the base of jets in different types of accreting black holes, and neutron stars.

It is not yet clear whether the magnetic field configuration in jets changes as a function of luminosity. Jets in quiescent LMXBs are weaker than the brighter hard state jets, their electron populations have lower maximum energies, and they may be dominated by partially non-relativistic, perhaps Maxwellian energy distributions, leading to steeper OIR SEDs than are typical for optically thin synchrotron emission \citep{shahet13,market15,plotet16}. So far, no intrinsic optical/infrared synchrotron polarization has been detected in any quiescent LMXB \citep{dubuch06,russfe08,baglet14a}, but in possibly all objects tested so far, other components such as the companion star and accretion disc are likely to dominate the emission in these systems in quiescence. In the radio regime, just one linear polarization constraint exists of the quiescent radio emission, of $< 2.11$ per cent ($3~\sigma$ upper limit) in V404 Cyg \citep{ranaet16}.

Swift J1357.2--0933 and A0620--00 are two black hole LMXBs which have reported spectral evidence for synchrotron emission from quiescent jets \citep{gallet07,fronet11,shahet13,russet13,plotet16,yang16}. Here we present infrared polarimetric observations of these two sources.

\section{Observations and data reduction}

\subsection{Infrared polarization observations}

Our targets were observed with the Long-slit Intermediate Resolution Infrared Spectrograph (LIRIS) in imaging polarimetry mode on the 4.2-m William Herschel Telescope (WHT) at the Observatorio del Roque de los Muchachos, La Palma, Spain on 2013 Feb 20 and 21 UT. A log of the observations is provided in Table 1. The LIRIS detector has $1024 \times 1024$ pixels, covering a field of view of $4.27^{\prime} \times 4.27^{\prime}$ with a pixel scale of $0.25^{\prime\prime}$ pixel$^{-1}$. Conditions were good, with the seeing ranging from $\sim 0.7$--1.2 arcsec. All exposures were made in a five-point dither pattern, with exposure times varying depending on the filter (see Table 1). For Swift J1357.2--0933, long repeats were made in $J$ and $K_{\rm S}$-band filters to obtain reasonable signal-to-noise ratio (S/N) for this fainter target, whereas shorter $Z$, $J$, $H$ and $K_{\rm S}$ sequences were carried out for the brighter source A0620--00. For each exposure four simultaneous images were obtained, one at each of the four polarization angles; 0$^{\circ}$, 45$^{\circ}$, 90$^{\circ}$ and 135$^{\circ}$ of the Wollaston prism. Half of the observations were made with the telescope rotator at $0^{\circ}$ and half at $90^{\circ}$ (except when the S/N was slightly higher in one camera orientation, where more images were acquired at the angle with lower S/N), in order to correct for the relative transmission factors of the ordinary and extraordinary rays for each Wollaston \citep[see][]{alveet11,zapaet11}.

The data reduction was performed using the \textsc{lirisdr} package developed by the LIRIS team in the \textsc{iraf} environment\footnote{IRAF is distributed by the National Optical Astronomy Observatory, which is operated by the Association of Universities for Research in Astronomy, Inc., under cooperative agreement with the National Science Foundation.} \citep[for details, see][]{alveet11}. Repeated sequences were combined and aperture photometry was then performed on the resulting images, and the normalized Stokes parameters $q$ and $u$, and the fractional linear polarization (FLP) and position angle (PA) were measured using equations (11--13) in \cite{alveet11}. Errors on FLP and PA were computed taking into account the errors associated with the raw counts at each polarization angle and polarization bias for low S/N data \citep{wardkr74} using the same method as \cite{russsh14}. The instrumental polarization is known to be very small for LIRIS; $< 0.1$ per cent \citep{alveet11}. A polarized standard star HD38563C was observed in all four filters to check for an offset in position angle. The PA measured agreed with literature values \citep{whitet92} in all filters to within 2--$12^{\circ}$ (errors were $< 1^{\circ}$; smaller than the literature values) with no systematic shift between the LIRIS and literature values, so no PA correction is applied to the data.

Short-term flux variability was also tested for using aperture photometry on shorter bins of the data. Using field stars from the Two Micron All Sky Survey \citep[2MASS;][]{skruet06}, we measured magnitudes and the rms variability as a fraction of the flux in each filter for the targets and six field stars close to the targets, except $Z$-band for A0620--00 in which the observations were too short to measure the rms, but approximate flux calibration was achieved using $z^{\prime}$-band SDSS magnitudes of field stars.

\begin{table}
\begin{center}
\caption{Log of observations of the two LMXB targets. The data were taken with the William Herschel Telescope (WHT), the Faulkes Telescope North (FTN) and the Faulkes Telescope South (FTS). For the WHT data a 5-point dither pattern was used; the total on source exposure time per row in the table is the individual exposure times (fourth column) $\times$ 5 (due to the dither pattern) $\times$ the number of repeats (fifth column).
}
\vspace{-2mm}
\begin{tabular}{lllll}
\hline
Telescope&UT Date 2013&Filter&Exp.&Dither \\
         &            &      &time&repeats\\
\hline
\multicolumn{5}{c}{--- A0620--00 ---} \\
FTN & 19 Feb 07:10--07:12 &$i^{\prime}$&100 s & -- \\
FTN & 19 Feb 07:12--07:14 & $V$       & 100 s & -- \\
FTN & 19 Feb 07:14--07:16 & $R$       & 100 s & -- \\
FTN & 19 Feb 10:04--10:12 &$i^{\prime}$&$4 \times 100$ s & -- \\
WHT & 20 Feb 21:02--21:18 & $J$	      & 60 s & 3  \\
WHT & 20 Feb 21:21--21:49 & $K_{\rm S}$ & 20 s & 11 \\
WHT & 20 Feb 21:51--22:15 & $H$	      & 30 s & 7  \\
WHT & 20 Feb 22:16--22:21 & $Z$	      & 60 s & 1  \\
WHT & 20 Feb 22:27--22:44 & $J$	      & 60 s & 3  \\
WHT & 20 Feb 22:46--23:14 & $K_{\rm S}$ & 20 s & 11 \\
WHT & 20 Feb 23:16--23:39 & $H$	      & 30 s & 7  \\
WHT & 20 Feb 23:40--23:52 & $Z$	      & 60 s & 2  \\
WHT & 20 Feb 23:56--00:13 & $J$	      & 60 s & 3  \\
WHT & 21 Feb 00:14--00:30 & $K_{\rm S}$ & 20 s & 6  \\
WHT & 21 Feb 00:31--00:59 & $H$	      & 30 s & 6  \\
WHT & 21 Feb 01:00--01:16 & $J$	      & 60 s & 3  \\
\vspace{2mm}
WHT & 21 Feb 01:20--01:35 & $K_{\rm S}$ & 20 s & 6  \\
\multicolumn{5}{c}{--- Swift J1357.2--0933 ---} \\
FTS & 18 Feb 18:09--18:13 & $I$       & 200 s & -- \\
WHT & 20 Feb 01:47--06:01 & $J$	      & 60 s & 40 \\
WHT & 20 Feb 06:05--06:25 & $K_{\rm S}$ & 20 s & 8  \\
WHT & 21 Feb 01:44--05:47 & $J$	      & 60 s & 40 \\
WHT & 21 Feb 05:49--06:45 & $K_{\rm S}$ & 20 s & 20 \\
\hline
\end{tabular}
\normalsize
\end{center}
\end{table}

\subsection{Optical observations}

Both LMXBs were observed at optical wavelengths within a few days of the infrared polarization observations using the 2-m Faulkes Telescopes, as part of an ongoing monitoring campaign of $\sim 40$ low-mass X-ray binaries \citep{lewiet08}. Swift J1357.2--0933 was observed in the Bessell $I$-band on 2013-02-18 using the Faulkes Telescope South (FTS) at Siding Spring, Australia (see Table 1).  Conditions were good and the seeing was 1.5$^{\prime\prime}$. The LMXB was faint but detected in the image. Flux calibration was achieved using the SDSS magnitudes of stars in the field, converting the SDSS magnitudes of the stars to Bessel $I$ magnitudes adopting the colour transformations in Table 3 of \cite*{jordet06}. The magnitude of Swift J1357.2--0933 was found to be $I = 20.27 \pm 0.25$.

A0620--00 was observed in Bessell $V$ and $R$-bands and four times in SDSS $i^{\prime}$-band on 2013-02-19 with the Faulkes Telescope North (FTN) located at Haleakala on Maui, Hawaii, USA (see Table 1). The conditions were worse than for Swift J1357.2--0933, with seeing of 2.1--2.4$^{\prime\prime}$, deteriorating to 3.5$^{\prime\prime}$ in the last $i^{\prime}$-band image. Despite the poorer conditions, A0620-00 is brighter and was detected in all images. SDSS magnitudes of three field stars were used for flux calibration. For $i^{\prime}$-band the SDSS magnitudes were used directly; for $V$ and $R$ conversions from \citet[][ table 7]{smitet02} were adopted \citep[no suitable conversions for $V$ and $R$ exist in][]{jordet06}. Each magnitude of A0620-00 in the four $i^{\prime}$-band exposures was consistent to within $1.1~\sigma$ of the mean value, so intrinsic variability was not detected. The average magnitude of A0620-00 was $V = 18.16 \pm 0.14$; $R = 17.17 \pm 0.05$; $i^{\prime} = 17.59 \pm 0.05$.

\section{Results and analysis}

\begin{figure}
\centering
\includegraphics[width=8.3cm,angle=270]{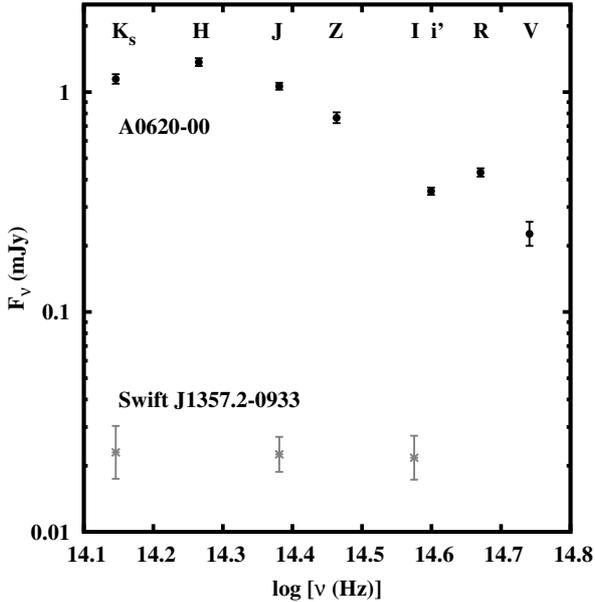}
\caption{Mean flux densities of the two LMXBs from optical to infrared in February 2013 (see Table 1 for dates of the observations).}
\end{figure}

\subsection{Flux and variability}

\begin{figure}
\centering
\includegraphics[height=8.3cm,angle=270]{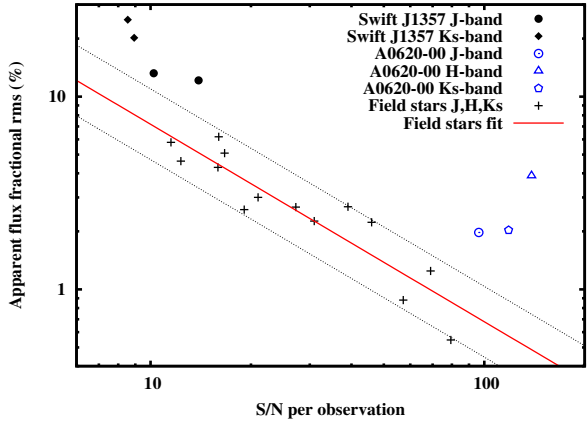}
\caption{Measured fractional rms variability as a function of the mean S/N in each image for the sources in the field of the two LMXBs. The fit to the field stars (red solid line) has been subtracted in quadrature from the measured rms of the LMXBs to estimate the intrinsic rms of the LMXBs (which are not plotted here, but are given in Table 2).}
\end{figure}

The de-reddened OIR spectral energy distributions (SEDs) of the two LMXBs are shown in Fig. 1. When more than one flux measurement was made in a particular filter, the mean magnitude was calculated from all observations, and the error represents the standard deviation of the magnitudes. The data were de-reddened using extinction values of $A_v = 1.05$ for A0620--00 and $A_v = 0.124$ for Swift J1357.2--0933 \citep{cantet10,armaet13,corret16}, adopting the extinction law of \citet*{cardet89}. We note that using other dust extinction laws in the infrared \citep{indeet05,chiati06} would change the $J$, $H$, $K_{\rm S}$ fluxes by $< 3$ per cent and $< 0.3$ per cent for A0620--00 and Swift J1357.2--0933 respectively, and so do not significantly alter their spectrum in Fig. 1.

Light curves of the two LMXBs and the six aforementioned (Section 2.1) close field stars were produced from each of the individual exposures, to investigate the intrinsic variability of the sources. The fractional rms variability amplitudes for each source in each filter on each date were calculated using equation (1) of \citet{gandet10}, which gives the excess variation above Poisson noise. The measured fractional rms variability is shown as a function of the S/N in Fig. 2. Many of the field stars, and Swift J1357.2--0933, have low S/N in each image, and there is a relation between apparent (measured) rms variability and S/N. The field stars are not expected to be variable and the relation shows the rms at low S/N is clearly not intrinsic variability of the source. We remove this contribution to the rms from the measurements of the two LMXBs, by subtracting the rms from the fit (solid line in Fig. 2) in quadrature, and the error is estimated from the error on the normalization of the fit (dotted lines in Fig. 2). The resulting rms values represent the intrinsic rms variability of the source, and are given in Table 2. Due to the exposure times, dithering pattern, cycling through the filters and total observation lengths, the time resolution and frequency range probed differed between filters and source (Table 2). The time resolution represents the average time between images on each date per filter, even when there were gaps between runs of consecutive images. The minimum and maximum values in the frequency range probed correspond to the total length of the observations and the smallest time between consecutive exposures, respectively. The results are discussed below.
\\
\\
\noindent \emph{A0620--00:} The mean flux density of A0620--00 (black circles in Fig. 1) is higher in the near-infrared (NIR) than the optical, and the SED is similar to that seen by \citet{gallet07} and \citet{fronet11}, suggesting that the curved spectrum is dominated by the companion star, which appears to peak (in flux density) near the $H$-band. The elevated data point at $\log \nu /$Hz $= 14.67$ ($R$-band) is most likely affected by the H$\alpha$ emission line from the accretion disc, a line often prominent in quiescent LMXBs \citep[e.g.][]{casach92,fendet09,casa15}. It was found from long-term OIR monitoring of A0620--00 \citep{cantet08} that the quiescent emission can be described by three states: passive (faint, in which the companion dominates), active (in which the optical emission is brighter by up to 0.5 mag and slightly bluer, likely due to the accretion disc), and loop (an intermediate case in which the source exhibits loops in its $V$--$I$ colour-magnitude diagram). The flaring behaviour affects not just the optical, but the NIR too, and NIR spectra also support a non-stellar contribution at these longer wavelengths \citep{fronet07}. Comparing the mean magnitudes we measure of $V = 18.16 \pm 0.14$ and $H = 14.76 \pm 0.05$ to Fig. 2 of \citet{cantet08}, both magnitudes are consistent with all three states, within uncertainties. It is therefore not clear whether A0620--00 was in an `active quiescent' state on the dates of our observations, although the elevated $R$-band flux does suggest the presence of a disc producing H$\alpha$ emission.

\begin{table}
\begin{center}
\caption{Table of variability results. The intrinsic fractional rms variability takes into account Poisson noise and the relation with S/N (Fig. 2). $\Delta t$ is the mean time resolution of consecutive observations.}
\vspace{-2mm}
\begin{tabular}{lllll}
\hline
Night&Filter&Intrinsic&$\Delta t$&Frequency        \\
     &      &rms (\%) &(s)       &range probed (Hz)\\
\hline
\multicolumn{5}{c}{--- A0620--00 ---} \\
\vspace{0.5mm}
2 & $J$         & $1.84^{+0.08}_{-0.19}$ & $1365$ & $6.7\times 10^{-5}$--$2.9\times 10^{-3}$\\
\vspace{0.5mm}
2 & $H$         & $3.86^{+0.02}_{-0.04}$ & $1226$ & $9.1\times 10^{-5}$--$5.0\times 10^{-3}$\\
\vspace{2mm}
2 & $K_{\rm S}$ & $1.95^{+0.05}_{-0.12}$ & $1402$ & $6.7\times 10^{-5}$--$6.6\times 10^{-3}$\\
\multicolumn{5}{c}{--- Swift J1357.2--0933 ---} \\
\vspace{0.5mm}
1 & $J$         & $11.2^{+1.2}_{-3.5}$ & $756$ & $8.8\times 10^{-5}$--$1.4\times 10^{-3}$\\
\vspace{0.5mm}
2 & $J$         & $11.0^{+0.7}_{-1.7}$ & $732$ & $7.2\times 10^{-5}$--$1.4\times 10^{-3}$\\
\vspace{0.5mm}
1 & $K_{\rm S}$ & $23.6^{+0.8}_{-2.1}$ & $301$ & $1.1\times 10^{-3}$--$3.3\times 10^{-3}$\\
2 & $K_{\rm S}$ & $18.5^{+1.0}_{-2.5}$ & $342$ & $3.2\times 10^{-4}$--$3.3\times 10^{-3}$\\
\hline
\end{tabular}
\normalsize
\end{center}
\end{table}

At longer wavelengths, a mid-IR excess above the companion emission, at a similar flux density to the radio emission, was interpreted as either dust from a circumbinary disc around the A0620--00 system, or synchrotron emission from quiescent jets in the system \citep{munoma06,gallet07}. The non-stellar IR emission was found to be flat and variable \citep{maitet11}, supporting the jet interpretation, and the spectral break in the jet spectrum was identified \citep{russet13} at a frequency of $(1.3 \pm 0.5) \times 10^{14}$ Hz (or 1.7--3.6 $\mu$m). \cite{fronet11} found that an upturn in the SED in A0620--00 in the far-UV could be explained by the extension of the optically thin synchrotron power law and is consistent with the jet model of \citet{gallet07}. Recently, multiple frequency observations carried out with the Karl G. Jansky Very Large Array in 2013 have revealed a highly inverted radio spectrum, with $\alpha \sim +0.7$ (where $F_{\nu} \propto \nu^{\alpha}$) and long-term variability of the radio emission \citep[Dincer et al. in preparation; see also][]{macdet15}. This is not unreasonable; a jet spectral index consistent with being this steep has been observed during outbursts in XTE J1118+480 and MAXI J1836--194, and in the latter source the radio jet was spatially resolved \citep{fendet01,russet15}. If this steep spectrum is not due to low frequency self-absorption, and instead extends to higher frequencies, it would have to either curve towards being flat \citep[curvature is seen in some sources, such as V404 Cyg;][]{russet13}, or the spectral break would have to reside at much lower frequencies than the NIR, in order for the NIR spectrum to be optically thin synchrotron. The infrared data in the literature imply a jet flux density of $\sim$0.1--0.4 mJy at NIR wavelengths \citep{gallet07,russet13}, which would be $\sim$7--40 per cent of the observed (total) flux density in the $J$--$K_{\rm S}$ wave-bands.
It is therefore possible that a polarization signal could be detected if the jet component is intrinsically polarized. 

The fractional rms variability of A0620--00 is clearly above the expected value for a zero intrinsic variability source at its S/N (Fig. 2), in all three wave-bands. The intrinsic rms variability is $\sim 1.8$--3.9 per cent (Table 2), with the highest rms in the $H$-band. On these timescales (the time resolution is $\sim 20$ minutes), the disc (if irradiated) and the jet could be variable. In addition, the observing period in total is 4.5 hours, which covers $\sim 60$ per cent of the orbital period of 7.75 hours, so the modulation of the companion is expected to be present. Inspecting the individual light curves, the mean flux level appears to stay roughly constant in the first 3.5 hours, then an increase of amplitude $\sim 0.1$ mag appears in the last 1 hour of the 4.5-hour observing period in all three wave-bands. The majority of the variability could be simply due to the orbital modulation, but we cannot rule out the presence of low-amplitude flares.

Previous optical variability studies \citep[e.g.][ and references therein]{shahet04} have revealed optical flares in A0620--00 lasting several minutes, above the orbital modulation. These flares have a spectral index of $\alpha \sim -1.4$, which are steeply red and similar to that measured for the variability in Swift J1357.2--0933 \citep{shahet13}, the latter of which was interpreted as synchrotron emission from the jet. Although faint in quiescence, one radio flare lasting $\sim 3$ minutes was observed from A0620--00 in quiescence \citep{palch04}, suggesting that there is indeed jet variability in quiescence in this source. This has been confirmed by Dincer et al. \citep[in preparation; see also][]{macdet15} who found long-term variability in the radio emission. If this variability dominates our intrinsic rms measurements, the lower fractional rms in $K_{\rm S}$-band compared to $H$-band (Table 2) suggests that the $\alpha \sim -1.4$ spectrum of the flares may peak around $H$-band (the flux density is also lower in $K_{\rm S}$-band), although the NIR variability measured here does not probe the shortest timescales of the flares observed by \citet{shahet04}. The decrease of rms variability at the longest wavelength ($K_{\rm S}$-band) coincides with the frequency of the jet spectral break at $\sim 1.3 \times 10^{14}$ Hz \citep{russet13}, which could be consistent since synchrotron-emitting regions overlap below the break, reducing the rms variability \citep[e.g.][]{malz14}. Alternatively, the orbital modulation of the companion could be producing these variations in flux, since the observing time spans more than half an orbital period. No obvious modulation is evident in the light curves, and it is not clear why there would be more fractional rms in $H$-band than $J$ and $K_{\rm S}$, but nevertheless the companion could be contributing to this variability.
\\
\\
\noindent \emph{Swift J1357.2--0933:} The mean flux density of Swift J1357.2--0933 appears to be constant from $K_{\rm S}$-band to $I$-band (Fig. 1), with the errors giving a spectral index of $\alpha = -0.06 \pm 0.51$. This spectral index is shallower than the OIR index $\alpha = -1.4 \pm 0.1$ observed by \citet{shahet13}, but is similar to that reported in the same wavelength range by \citet{plotet16}. The $I$-band data were taken 1.5 days before the (night 1) NIR data (Table 1), and the mean flux is known to vary on timescales shorter than this, so caution should be made when joining the $I$-band data in the SED. Nevertheless, in the context of the jet synchrotron interpretation of \citet{shahet13} and \citet{plotet16}, the data imply the optically thick, flat spectrum jet may have dominated from $K_{\rm S}$-band to $I$-band around the dates of our observations, with the jet spectral break residing at a higher frequency than $I$-band; $\nu_{b} \geq 3.8\times 10^{14}$ Hz \citep[][ found $\nu_{b} \approx 2$--$5\times 10^{14}$ Hz]{plotet16}.

The intrinsic fractional rms variability in $K_{\rm S}$-band is high; $\sim 18$--24 per cent on both dates, similar to that found previously \citep{shahet13}. In $J$-band, the rms was $\sim 11$ per cent on both dates, lower than that found by \citet{shahet13}. The resulting spectral index of the rms itself is $\alpha \sim -1.3$, which is similar to that found by \citet[][ see their Fig. 4, lower panel]{shahet13}. This implies that perhaps the underlying variable jet component is optically thin, even though the total observed spectrum is flat. Alternatively, the optically thick, flat spectrum synchrotron emission from the jet could produce the total spectrum and variability, but the longer wavelengths have higher rms variability on these timescales for some as yet unknown reason. One possibility is that the power density spectrum of the variability is expected to peak at higher frequencies (i.e., the variations are faster) at shorter wavelengths \citep[as expected by models, e.g.][]{malz14}, and we are here probing slower timescales than this, in which case on these slow timescales the longer wavelengths may have more rms variability power than the shorter wavelengths, as observed between $K_{\rm S}$ and $J$-bands. This may be consistent with the results of \citet{gandet11}, who found that in GX 339--4, at wavelengths close to the jet spectral break the rms variability was of high amplitude, but the variations were slow, with smooth changes on timescales of minutes--hours. We therefore consider both options, optically thin and optically thick synchrotron, are plausible possibilities for the origin of the NIR spectrum and variability.

\subsection{Polarization}

\begin{table}
\begin{center}
\caption{Table of polarization results.}
\vspace{-2mm}
\begin{tabular}{lllll}
\hline
Night&Filter&FLP (\%)&PA ($^{\circ}$)&$\sigma_{\rm FLP}^a$\\
\hline
\multicolumn{5}{c}{--- A0620--00 ---} \\
2 & $Z$         & $1.42 \pm 0.63$ & $169.6 \pm 11.6$ & 2.26 \\
2 & $J$         & $1.37 \pm 0.21$ & $151.1 \pm 4.3$  & 6.58 \\
2 & $H$         & $1.20 \pm 0.20$ & $149.7 \pm 4.7$  & 6.04 \\
\vspace{2mm}
2 & $K_{\rm S}$ & $0.92 \pm 0.28$ & $126.7 \pm 8.3$  & 3.30 \\
\multicolumn{5}{c}{--- Swift J1357.2--0933 ---} \\
1 & $J$         & $6.84 \pm 4.17$ & $18.2 \pm 14.9$  & 1.64 \\
2 & $J$         & $3.16^{+3.60}_{-3.16}$ & $38.0 \pm 21.5$ & 0.88 \\
2 &$K_{\rm S}^b$& $6.41 \pm 5.65$ & $159.0 \pm 18.9$ & 1.13 \\
1+2&$J$         & $7.27 \pm 2.74$ & $25.4 \pm 10.1$  & 2.65 \\
1+2&$K_{\rm S}^b$&$8.44 \pm 4.76$ & $0.4 \pm 14.1$   & 1.77 \\
\vspace{2mm}
1+2&$J+K_{\rm S}$&$8.00 \pm 2.48$ & $14.6 \pm 8.5$   & 3.22 \\
\multicolumn{5}{c}{--- Field star$^c$ close to Swift J1357.2--0933 ---} \\
1+2&$J+K_{\rm S}$&$0.54^{+1.12}_{-0.54}$ & $75.2 \pm 32.7$  & 0.48 \\
\hline
\end{tabular}
\normalsize
\end{center}
$^a$The significance of the measured polarization is given, $\sigma_{\rm FLP} = $ FLP $/$ $\Delta $FLP.
$^b$The polarization of Swift J1357.2--0933 could not be measured on night 1 in the $K_{\rm S}$-band because only one of the two camera rotation angles were used. These data were included in the night 1+2 combined measurement.
$^c$A star in the field of Swift J1357.2--0933 close to (and slightly brighter than) the LMXB is shown for comparison. It is found to be unpolarized, indicating there is no systematic polarization introduced by combining both nights of data.
\end{table}

\begin{figure}
\centering
\includegraphics[height=8.3cm,angle=270]{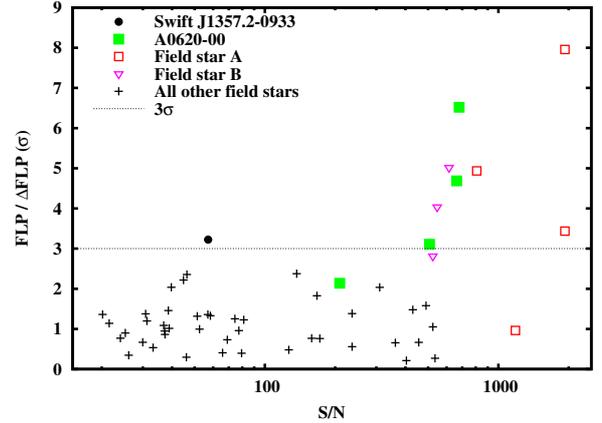}
\caption{The confidence of measured polarization (FLP divided by its error) is shown as a function of the S/N of the observation. The horizontal dotted line indicates a detection level of polarization at the 3 $\sigma$ level.}
\end{figure}

\begin{figure}
\centering
\includegraphics[width=8.3cm,angle=0]{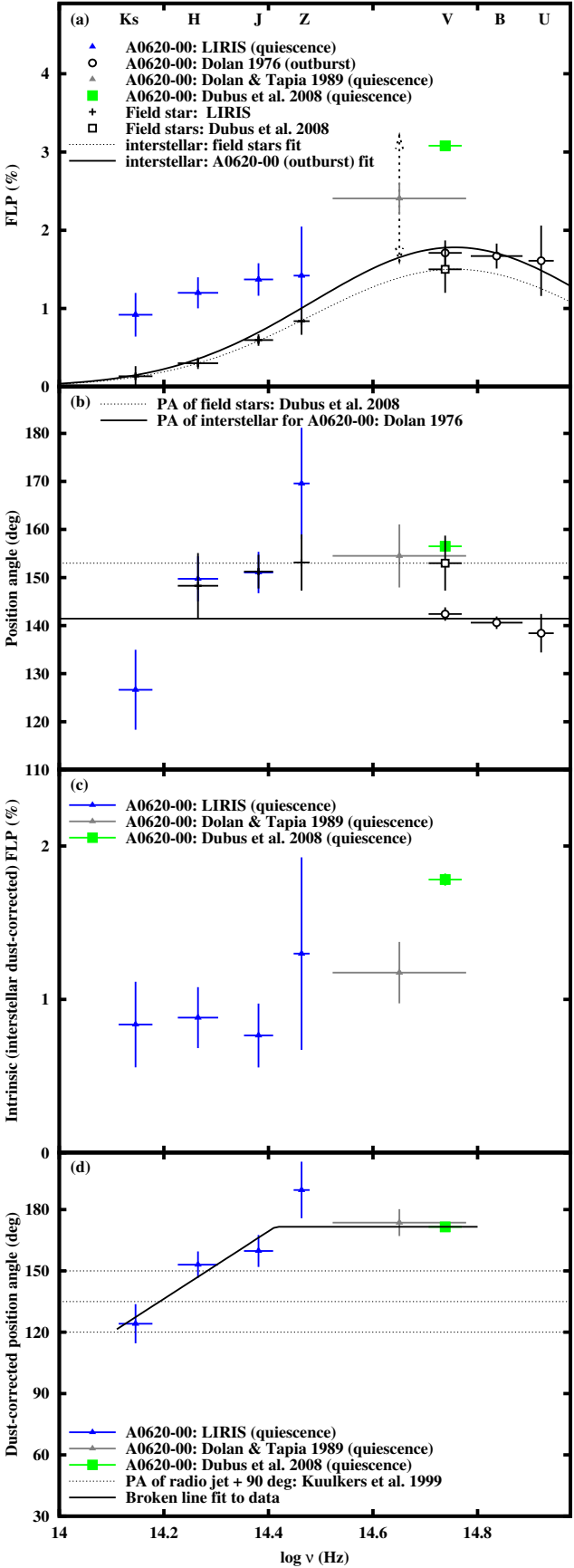}
\caption{FLP and PA spectra of A0620--00 and field stars (see text).}
\end{figure}

The polarization results for the two LMXBs in each filter are given in Table 3. For A0620--00, FLP is detected at a confidence level of $> 6~ \sigma$ in $J$ and $H$-bands, $3.3~\sigma$ in $K_{\rm S}$-band and $2.3~\sigma$ in $Z$-band. The level of FLP decreases with wavelength, from $\sim 1.4$ per cent in $Z$-band to $\sim 0.9$ per cent in $K_{\rm S}$-band. For Swift J1357.2--0933, no significant FLP is detected on either of the two nights of data in each wave-band, with the most significant being FLP $= 6.8 \pm 4.2$ per cent in $J$-band on night 1, a confidence of detection of FLP of only $1.6~\sigma$. Too few $K_{\rm S}$-band data were taken on night 1, the S/N was low, and it was not possible to measure the FLP in this wave-band on night 1. However, by combining the data of night 1 and night 2, we measure FLP $= 7.3 \pm 2.7$ per cent in $J$-band averaged over both nights, and FLP $= 8.4 \pm 4.8$ per cent in $K_{\rm S}$-band. Finally, we combine the $J$ and $K_{\rm S}$-band data together over both nights to achieve one, high S/N measurement of the FLP, resulting in FLP $= 8.0 \pm 2.5$ per cent, which is a detection of intrinsic polarization at the $3.2~\sigma$ level.

To further assess the significance of this result, we also measured the FLP of all field stars in each filter, that were bright enough and visible in the images at both camera rotation angles. The fields of the two LMXBs, and another LMXB 4U 0614+09 (results for this source are in preparation and will be published elsewhere) were used. Out of $>50$ measurements of FLP from the three fields, only A0620--00, Swift J1357.2--0933 and the two brightest field stars had significant ($>3~\sigma$) detections of FLP above zero (Fig. 3). These two field stars have FLP that follows the expected relation with wavelength for interstellar dust (see Fig. 4). The only other star in the field of Swift J1357.2--0933 had a FLP, measured in the combined night 1 and 2 $J$ and $K_{\rm S}$ data, of $0.5^{+1.1}_{-0.5}$ per cent (a significance of $<1~\sigma$; bottom row of Table 3), which further suggests that the FLP measured for Swift J1357.2--0933 is real and intrinsic to the source, although attention should be made to the significance of the detection ($3.2~\sigma$).
\\
\\
\noindent \emph{A0620--00:} To investigate the origin of the polarization in A0620--00, we compare our NIR measurements with optical FLP values from the literature. \citet{dolata89} found that the optical (white light) FLP varied with the orbital phase in quiescence, with values of FLP $\sim 1.6$--3.2 per cent. \citet{dubuet08} also measured FLP $= 3.08 \pm 0.04$ per cent in $V$-band during quiescence but did not detect the orbital modulation. One reason could be that the source may have been in a different (passive/loop/active) state \citep{cantet08}. In Fig. 4, the FLP (panel (a)) and PA (panel (b)) of A0620--00 are shown as a function of frequency from optical (literature data) to NIR (our LIRIS data). The phase-averaged quiescent optical FLP \citep[from][]{dolata89,dubuet08} are shown, with the dotted lines with arrows indicating the range of FLP values measured over the orbital period by \citet{dolata89}. Optical FLP measurements from the 1975 outburst \citep{dola76} are also included, as are NIR (LIRIS) and optical \citep{dubuet08} FLP measurements of field stars close to A0620--00, for comparison. The field stars are polarized due to the preferential orientation of dust grains between the stars and the Earth, and the FLP as a function of frequency can be well fit by the expected empirical relation \citep{serk73} for FLP due to interstellar dust (see dotted line in the upper panel of Fig. 4). The optical and NIR PA of the polarization of the field stars is found to agree, with PA $\sim 153^{\circ}$ (dotted line in second panel). The optical outburst FLP values of A0620--00 lie close to (but are slightly higher than) the relation fit to the field stars for interstellar dust. We fit the same dust relation to these optical data, changing only the peak FLP, not the frequency of the peak (solid curve in Fig. 4a). The PA of this is found to be $\sim 141^{\circ}$, slightly different to that of the field stars (Fig. 4b). The quiescent optical \citep{dolata89,dubuet08} and NIR (LIRIS) FLP clearly lie above the relation for interstellar dust, and so must have a different origin.

We remove the contribution of the dust to the observed FLP and plot the remaining FLP in Fig. 4(c). This is achieved using the best fit relation for dust polarization to the optical data of A0620--00, and subtracting its wavelength-dependent Stokes $q$ and $u$ values from the observed $q$ and $u$ for each measurement. The FLP and PA of the non-dust (i.e. intrinsic) data are then calculated from these values of $q$ and $u$. The dust-corrected FLP of A0620--00 is $\sim 1$--2 per cent at optical and $\sim 0.5$--1.0 per cent at NIR wavelengths. The optical FLP is likely to be due to Thomson scattering \citep[e.g.][]{browet78,bochet79} within the binary system, since it has been seen to vary on the orbital period \citep{dolata89}. For such scattering the FLP is expected to be either constant with wavelength, or slightly decreasing from optical towards NIR wavelengths \citep[e.g.][]{blaeag96,schuet04}, and the PA should be constant.
The clear decrease of FLP from optical to NIR (Fig. 4c) could therefore perhaps be explained by scattering. In the $Z$ and $J$-bands, the PA is consistent (within $2~\sigma$) with that of the optical (Fig. 4d). The PA in $H$ and $K_{\rm S}$-band differ from the optical PA by $3.0~\sigma$ and $5.1~\sigma$ respectively, and a broken line fit (Fig. 4d) describes well the change in PA. Scattering is therefore a plausible origin to the optical and some of the NIR polarization, and can account for the FLP and PA in all bands except the $H$ and $K_{\rm S}$-bands.
Since the optical FLP can vary with phase of the orbit \citep{dolata89}, one possible reason that the NIR FLP is lower than the optical FLP is that the NIR data could have been acquired at a specific phase of the orbit. The optical quiescent data of A0620--00 in Fig. 4 are the orbital phase-averaged values of FLP. Our NIR data span 4.5 hours, which is $\sim 60$ per cent of the orbital period, which is a substantial fraction, so it is possible, but unlikely, that our NIR data have a lower FLP because of the phase dependency and the orbital phase of the observations. In addition, the data in all NIR wave-bands span the same orbital phase, yet the $K_{\rm S}$-band PA differs significantly to the $J$ and $H$-band PA. Even if phase dependency is not the reason for the low NIR FLP, scattering could explain the polarization in the $J$ to $V$-bands.

While the FLP in $H$ and $K_{\rm S}$-bands is similar to that in $J$ and $Z$, the PA is not. The $K_{\rm S}$-band PA is $124.1 \pm 9.6^{\circ}$, which differs by $\sim 48^{\circ}$ to the PA of the optical bands. If the process is scattering, the PA is not expected to change with wavelength, so this $5.1~\sigma$ discrepancy is significant and we have to investigate other sources of polarization.
By assuming that the scattering would produce a PA in $K_{\rm S}$-band that is the same as at optical wavelengths ($\sim 173^{\circ}$) and a FLP that follows the trend of FLP with wavelength (i.e. about the same FLP as is shown in Fig. 4c), we 
estimate the remaining polarization (dust-corrected and scattering-corrected) in $K_{\rm S}$-band is FLP $\sim 1.25 \pm 0.28$ per cent.

A possible origin is polarization due to synchrotron emission from jets in A0620--00. The jet does make a contribution to the NIR flux in A0620--00 (see Section 3.1); we estimate from the infrared data \citep{gallet07,russet13} that it contributes $\sim 8$ -- 37 per cent of the $K_{\rm S}$-band flux density. The position angle of the resolved radio jet imaged during its 1975 outburst was constrained by \citet{kuulet99} to be $45 \pm 15^{\circ}$. For synchrotron, the PA of the polarization traces the electric field vector in the emitting region, and the magnetic field vector differs from this by $90^{\circ}$. In Fig. 4(d), the angle perpendicular to the resolved radio jet axis, and its error, are shown as horizontal dotted lines. The measured polarization PA in $K_{\rm S}$-band and $H$-band are consistent within errors with this angle. 
If synchrotron emission from the jet dominates the $K_{\rm S}$-band polarization, then this implies the magnetic field vector is parallel to the jet axis. This is the same as found in other LMXBs from NIR data \citep{russfe08,russet11} and radio data \citep{curret15}. Whether the synchrotron is optically thin, or partially self-absorbed (i.e. the flat spectrum that extends from NIR to radio), the polarization will be dominated by the optically thin regions, which have PAs aligned perpendicular to the magnetic field \citep[e.g.][]{zdziet14,curret15}. The $K_{\rm S}$-band FLP is $\sim 1.25 \pm 0.28$ per cent and if this is produced by the jet which contributes $\sim 8$ -- 37 per cent of the flux density, this would imply the jet contribution itself is polarized at a level of FLP$_{\rm jet} \sim 3$--18 per cent, suggesting that the magnetic field in the jet is moderately tangled.
\\
\\
\noindent \emph{Swift J1357.2--0933:} The spectral and rapid variability properties of Swift J1357.2--0933 favour a synchrotron process producing the NIR emission in quiescence. This is therefore the most likely origin of the measured FLP of $8.0 \pm 2.5$ per cent. The line of sight extinction of $A_v = 0.124$ towards the source is way too low to produce this level of polarization due to dust at NIR wavelengths. There is a linear relation between the maximum optical polarization caused by dust extinction; FLP$_{\rm max} = 3 A_{\rm v}$ \citep{serket75}, which implies FLP$_{\rm dust} < 0.4$ per cent for Swift J1357.2--0933, and much lower in the NIR regime. Scattering on the disc surface also seems unlikely since the disc does not dominate the NIR emission, so the intrinsic polarization of the disc would have to be even higher than that measured, which is unlikely \citep[scattering usually produces a few per cent FLP in LMXBs; e.g.][]{dolata89,glioet98}. Emission lines from the accretion disc have been detected in quiescence \citep{torret15,mataet15}, but its contribution to the continuum is likely to be low, since the UV flux is much fainter than the optical flux \citep{plotet16}. The X-ray luminosity is also very low, possibly the lowest measured of all black hole LMXBs \citep{armaet14}. The polarization of $8.0 \pm 2.5$ per cent implies the magnetic field in the region of the jet near its base is moderately tangled. However, the FLP is fairly high for LMXB jets, being comparable to that measured in Cyg X--2 \citep{shahet08}, lower than 4U 0614+09 \citep[if the disc is unpolarized in that source;][]{baglet14b} and much lower than Cyg X--1 \citep{lauret11,russsh14,rodret15}, but higher than GX 339--4, Sco X--1, GRO J1655--40 and XTE J1550--564 \citep{shahet08,russfe08,russet11,chatet11}. The length of the observations implies a continuously launched (stable on long timescales), highly variable (on short timescales) jet in Swift J1357.2--0933, which has a moderately tangled magnetic field close to the jet base. Alternatively, the magnetic field could be highly ordered, but with a variable PA on short timescales, which would smear out the high polarization on longer timescales than the variability.

During outburst, the optical SED of Swift J1357.2--0933 was well approximated by a non-irradiated accretion disc \citep{armaet13,shahet13,wengzh15}. Optical dips were best explained by a torus obscuring the inner regions that moved outwards during the outburst \citep{corret13}. Radio emission from a jet was detected during its outburst \citep{sivaet11} and excess NIR emission could be explained by the flat radio jet spectrum extending to NIR wavelengths \citep{shahet13}. The radio jet was not resolved, so its orientation is unknown. However, if the NIR polarization we detect in quiescence is produced by the jet, the polarization PA will be parallel to the electric field vector. The magnetic field vector is perpendicular to this and is expected to be parallel to the jet axis \citep[this has been found empirically; see][ and above]{curret15}. We can therefore speculate that the jet axis has a PA of $75 \pm 33^{\circ} + 90^{\circ} = 165 \pm 33^{\circ}$ on the plane of the sky. This cannot be tested until a future outburst because the radio jet emission is extremely faint in quiescence \citep{plotet16}. Radio VLBI data could be taken during a future outburst to test this prediction, assuming there is no long-term change in the position angle of the jet, due to a misalignment of the jet with the disc axis.

\section{Conclusions}

NIR polarimetric observations of two black hole LMXBs in quiescence have been presented. In both sources, we find for the first time (at the $>3~\sigma$ level) the polarimetric signature of synchrotron emission, which we ascribe to the weak collimated outflows launched at low accretion rates in these systems. This further supports \citep[in addition to the previous spectral and timing evidence; e.g.][]{gallet07,shahet13,russet13,plotet16} the existence of jets launched in quiescence in black hole X-ray binaries. We constrain the level of polarization of the synchrotron emission to be $8.0 \pm 2.5$ per cent and $\sim 3$--18 per cent in Swift J1357.2--0933 and A0620--00, respectively. The latter result has a large range due to the uncertain NIR jet contribution, which we estimate to be $\sim 8$ -- 37 per cent of the observed $K_{\rm S}$-band flux.

For A0620--00 we identify two additional sources of line-of-sight polarization -- interstellar dust (which produces 1--2 per cent FLP at optical, and $\ll 1$ per cent at NIR wavelengths, respectively), and Thomson scattering within the system (1--2 per cent FLP at OIR wavelengths). In the $H$ and $K_{\rm S}$ wave-bands (the longest wavelengths), the position angle of the polarization changes significantly, which is due to an excess of FLP of $\sim 1.25 \pm 0.28$ per cent, which we argue is due to synchrotron emission which contributes a small amount of the flux at these wavelengths. The PA of this synchrotron component implies that the orientation of the magnetic field is approximately parallel to the axis of the radio jet of A0620--00 imaged during outburst, which is in line with the magnetic field orientations of other LMXBs at higher accretion rates. The jet of Swift J1357.2--0933 has not been spatially resolved. Nevertheless, if the same magnetic field orientation is true for this source, then our polarization results imply the jet axis has a position angle of $\sim 165 \pm 33^{\circ}$. In general, the measured polarization properties put in the context of previous radio and OIR results taken in quiescence, suggests that there is a continuously launched (on long timescales), highly variable (on short timescales) jet launched from these systems during the long periods of quiescence between outbursts.

In addition to the polarimetric results, intrinsic variability is detected in both sources on minute--hour timescales. The high fractional rms variability of 11--24 per cent in Swift J1357.2--0933 has a NIR spectral index of $\alpha \sim -1.3$, very similar to that measured by \citet{shahet13}, even though the observed OIR flux spectrum is flat ($\alpha \sim -0.1$), which is similar to \citet{plotet16}. This suggests that either the partially self-absorbed (optically thick) synchrotron dominates the emission and there is stronger variability on these long timescales at the longer wavelengths, or that the synchrotron emission is optically thin, and another component, for example the accretion disc, produces some of the shorter wavelength flux. The variability is much weaker in A0620--00; the NIR fractional rms of $\sim 2$--4 per cent could be explained by accretion activity and/or the orbital modulation of the companion star.

The polarization results imply that the magnetic fields near the base of black hole LMXB jets are moderately tangled, with a preferential orientation of the magnetic field along the jet axis. This is similar to the properties derived from OIR polarization studies of LMXBs at higher accretion rates, but not for all cases. Future observations, especially monitoring changes of the magnetic field properties throughout an outburst of a transient LMXB, could probe for the first time how the jet properties depend on accretion rate/state. In addition, the magnetic fields in neutron star jets have not yet been studied at low accretion rates, so future observations of a quiescent neutron star system could test the ubiquity of these tangled magnetic fields in LMXBs.

\section*{Acknowledgements}
We thank Jos\'e Acosta-Pulido and Antonio Pereyra for help with the LIRIS data reduction pipeline. We thank the pupils and teacher Pauline Hayes at Berlin International School for taking some of the Faulkes Telescope observations. Based on scheduled observations made with the William Herschel Telescope operated on the island of La Palma by the Isaac Newton Group in the Spanish Observatorio del Roque de Los Muchachos. The Faulkes Telescopes  are  maintained  and  operated  by  Las
Cumbres   Observatory   Global   Telescope   Network.
TS acknowledges support from the Spanish Ministry of Science and Innovation (MICINN) under the grant AYA2013--42627.

\bsp	% typesetting comment
\label{lastpage}
\end{document}